\documentclass[conference]{IEEEtran}
\IEEEoverridecommandlockouts

\usepackage{cite}
\usepackage{amsmath,amssymb,amsfonts}
\usepackage{algorithmic}
\usepackage{graphicx}
\usepackage{textcomp}
\usepackage{xcolor}
\usepackage{booktabs}
\usepackage{makecell}
\usepackage{tabularx}
\usepackage{newunicodechar}
\usepackage{multirow}
\usepackage[table]{xcolor} 
\def\BibTeX{{\rm B\kern-.05em{\sc i\kern-.025em b}\kern-.08em
    T\kern-.1667em\lower.7ex\hbox{E}\kern-.125emX}}
\begin{document}

\title{AVSRBench: A Multi-Condition AVSR Benchmark\\

\thanks{This publication emanates from research supported by Taighde Éireann – Research Ireland, Grant number 22/FFP-A/11059.}
}
\author{\IEEEauthorblockN{Rishabh Jain}
\IEEEauthorblockA{\textit{Sigmedia Group, School of Engineering} \\
\textit{Trinity College Dublin, Ireland}\\
rijain@tcd.ie}
\and
\IEEEauthorblockN{Naomi Harte}
\IEEEauthorblockA{\textit{Sigmedia Group, School of Engineering} \\
\textit{Trinity College Dublin, Ireland}\\
nharte@tcd.ie}
}
\maketitle

\begin{abstract}
While AVSR has achieved sub-1\% word error rates on the standard LRS3 benchmark, its reliance on broadcast speech obscures whether this reflects true generalization or just domain adaptation. To investigate this gap, we evaluate three AVSR architectures across six conditions: controlled broadcast speech, fixed-grammar utterances, hyper-articulated Lombard speech, read speech from professional lipspeakers and non-professional speakers, and spontaneous multi-party video conversations. We find that visual-only performance deteriorates rapidly beyond broadcast domains, and audio-video fusion mainly benefits Lombard speech environments. Visual understanding degrades sharply at 90° profile views, with multimodal systems relying largely on acoustic fallback. Additionally, speaker articulation proves more critical than minor camera shifts, and LLM-based architectures suffer from poor out-of-domain generalization. Our work highlights a significant generalization gap in current AVSR research. To address this, we also introduce RoomReader-AV as a new benchmark for AVSR and release a unified data preprocessing pipeline to make comprehensive multi-condition evaluation accessible.
\end{abstract}

\begin{IEEEkeywords}
AVSR, Lipreading, LRS Datasets, Multi-condition evaluation, Benchmarking AVSR, RoomReader
\end{IEEEkeywords}

\section{Introduction}

Audio-visual speech recognition (AVSR) has made major progress on standard benchmarks over the past several years \cite{sheng2024deep,kit2025phoneme,prajwal24_interspeech, avformer, ahn24_interspeech}. Models such as AV-HuBERT \cite{avhubert,shi22_interspeech}, Auto-AVSR \cite{autoavsr}, and Llama-AVSR \cite{llama_avsr} have achieved word error rates (WER) below 2\% on LRS2 \cite{lrs2} and LRS3 \cite{lrs3} broadcast corpora, the two datasets that dominate AVSR research. With performance approaching near-perfect on these controlled broadcast benchmarks, it is easy to get the impression that AVSR is close to a solved problem \cite{xia2020audiovisual}.

Some recent studies \cite{liu2023synthvsr, ahn24_interspeech, su2026m2savsrmodalityawaremultiviewselfsupervised,tian2026seeingcontextrichvisual} have attempted to move beyond controlled broadcast corpora by simulating more challenging conditions, such as cocktail-party noise \cite{cocktail_party, vi_cocktail,cocktail_benchmark}, in-the-wild speech \cite{Ma2022, lrs_voxmm, djilali2024vsr}, video conferencing \cite{av_conference}, and Lombard speech \cite{lombard_speech, lombard2, ma19b_interspeech}. Cocktail-party evaluations show that models can degrade sharply from 7\% to over 69\% WER when interfering speakers and background noise are present \cite{nguyen25b_interspeech}. In-the-wild benchmarks also report an average 30\% absolute WER spike across VSR models relative to LRS3, suggesting that broadcast-trained representations do not generalize well beyond controlled lip movements \cite{djilali2024vsr}. Video conferencing scenarios show a similar pattern, with Auto-AVSR’s audio-visual (AV) WER rising from under 1\% to over 33\% on Zoom \cite{av_conference}. The LRS-VoxMM benchmark \cite{lrs_voxmm} confirms that real-world audio distortions like reverberation make conversational speech significantly harder to transcribe than LRS3. Lombard speech \cite{mcgurk1976hearing} adds another challenge, since noise-induced hyperarticulation \cite{hyper} changes both acoustic and visual speech production in ways that broadcast-trained models are not typically exposed to \cite{ma19b_interspeech}. Even recent LLM-based decoders \cite{su2026robust,mms} mainly improve performance through lexical decoding, rather than better visual feature representation, which leaves the visual encoder as a major bottleneck \cite{jain2026hypeinsightrethinkinglarge}. This limitation stems from models relying heavily on training word frequencies rather than actual visual perception \cite{lip_reading_gap}. Lin et al. \cite{multivsr_lrs3} use MultiVSR \cite{multivsr} subsets to confirm that this vocabulary reliance impacts performance severely, even in LRS3-matching environments. More broadly, these studies show that AVSR performance is highly sensitive to out-of-domain conditions, but they still examine each factor in isolation \cite{lin2025uncovering, hong2023watch}. 

Most AVSR research still evaluates models on broadcast speech corpora, where speakers face the camera, lighting is good, and speech is clearly articulated. The LRS3 \cite{lrs3} test set is less than an hour, which can inflate performance and hide how sharply models degrade outside the broadcast setting. Many AVSR models ship with evaluation code that is tightly coupled to the LRS2 \cite{lrs2} and LRS3 benchmark datasets, and running them on new corpora often requires custom data preparation, cleaning, and decoding scripts that are not standardized or easy to reuse. This makes it hard to compare models fairly across datasets and increases the effort required to move beyond LRS2 and LRS3 in a systematic way.

To address these gaps, this paper makes three core contributions. First, we release a standardized audio-visual and lipreading data preparation toolkit designed for common evaluation across multiple datasets. This toolkit integrates with Auto-AVSR and AV-HuBERT codebases allowing researchers to streamline multi-condition evaluations. Second, we introduce RoomReader \cite{reverdy2022roomreader} as a benchmark for spontaneous, multi-party video conferencing conversations. It captures spontaneous Zoom interactions rather than controlled broadcast speech and thus provides a challenging setting for conversational AVSR. Third, we conduct a systematic evaluation of three modern AVSR systems across six publicly available datasets that have not previously been benchmarked together. This evaluation tests variables that earlier work often overlooks, including non-frontal camera angles, speaker articulation, Lombard speech, and video conferencing conditions.

Our evaluation under visual-only (VO), audio-only (AO), and audio-visual (AV) settings shows several clear patterns. VO recognition fails badly outside broadcast conditions, and neither more training data nor an LLM-based decoder solves this problem. AV fusion helps mainly in Lombard speech environments; otherwise, it offers little to no advantage over AO recognition. We also find that minor visual changes have little effect, but extreme profile views and speaker articulation matter much more. Our work shows that current AVSR systems remain brittle outside controlled broadcast data.

\section{Datasets and Processing}

\subsection{Dataset Overview}
We evaluate across six datasets spanning a range from broadcast speech to fully spontaneous conversation, as summarized in Table \ref{tab:dataset_summary}.

\begin{table}[t]
  \caption{Dataset Summary for Evaluation}
  \label{tab:dataset_summary}
  \centering
\renewcommand{\arraystretch}{1.15}
  \begin{tabular}{|l|c|c|c|l|}
    \hline
    \rowcolor{red!10!blue!15}
    \textbf{Dataset} & \textbf{Hrs.} & \textbf{Utter.} & \textbf{Spk.} & \textbf{Condition} \\
    \hline

    LRS2        & 0.80h  & 1,243  & - & Broadcast (BBC TV) \\
    \hline
    LRS3        & 0.84h  & 1,321  & -      & Broadcast (TED/TEDx) \\
    \hline
    GRID        & 23.0h  & 32,895 & 34       & Fixed-vocabulary syntax \\
    \hline
    LombardGrid  & 7.35h  & 10,749 & 54       & Lombard speech in noise \\
    \hline
    TCD-TIMIT   & 22.0h  & 13,826 & 62       & Phonetically rich read speech \\
    \hline
    RoomReader  & 6.49h  & 10,324 & 118      & Spontaneous conversational \\
    \hline

  \end{tabular}

  \vspace{4pt}
  {\footnotesize \textit{Hrs. = Hours, Utter. = Utterances; Spk. = Speakers.}}
\vspace{-1.5em}
\end{table}

\textbf{LRS2} \cite{lrs2} consists of clips from BBC television broadcasts. We use only the test set in this work. 

\textbf{LRS3} \cite{lrs3} consists of TED and TEDx talk clips. It is the primary training domain for AV-HuBERT and Auto-AVSR and serves as the main in-domain benchmark throughout this paper.

\textbf{GRID} \cite{Grid} is a 34-speaker corpus with a fixed six-word command grammar. The audio is clean and the recordings are frontal, but the vocabulary is unseen during training. This makes GRID a useful test of generalization beyond the training vocabulary and sentence structure.

\textbf{LombardGrid} \cite{lombardgrid} contains 54 speakers recorded while listening to 80 dB background noise, which elicits Lombard speech: louder voice, higher pitch, and more exaggerated lip movements. Crucially, synchronized frontal (0°) and profile (90°) cameras allow direct comparison of camera angle effects within the same recording session.

\textbf{TCD-TIMIT} \cite{harte2015tcd} contains 62 speakers recording 6,913 phonetically rich TIMIT sentences. The dataset is first divided by articulation expertise into three professional lipspeakers (trained to articulate clearly for hearing-impaired audiences) and 59 non-professional volunteers. Each of these groups is then further subdivided by camera angle into straightcam (0°) and 30degcam (30°) subsets.

\textbf{RoomReader} \cite{reverdy2022roomreader} contains 30 Zoom-based tutorial sessions with 118 participants and manually corrected transcriptions with word-level boundaries. Speech is fully spontaneous and conversational. The dataset provides distinct audio streams: a session-level audio file containing all participants' speech (which inherently includes overlapping speech from other speakers), and individual audio tracks containing each speaker's isolated audio. While our preprocessing pipeline supports both formats, we report results exclusively on the individual participant stream to evaluate noisy, single-speaker Zoom-based speech. For our evaluation, we further divide the data into Easy and Hard subsets, the specific criteria for which are detailed later in Section \ref{sec:rr_av}. RoomReader has not previously been used for AVSR evaluation and offers a novel and challenging out-of-domain benchmark for AVSR.

\begin{figure}[t]
\centering
\includegraphics[width=1\linewidth]{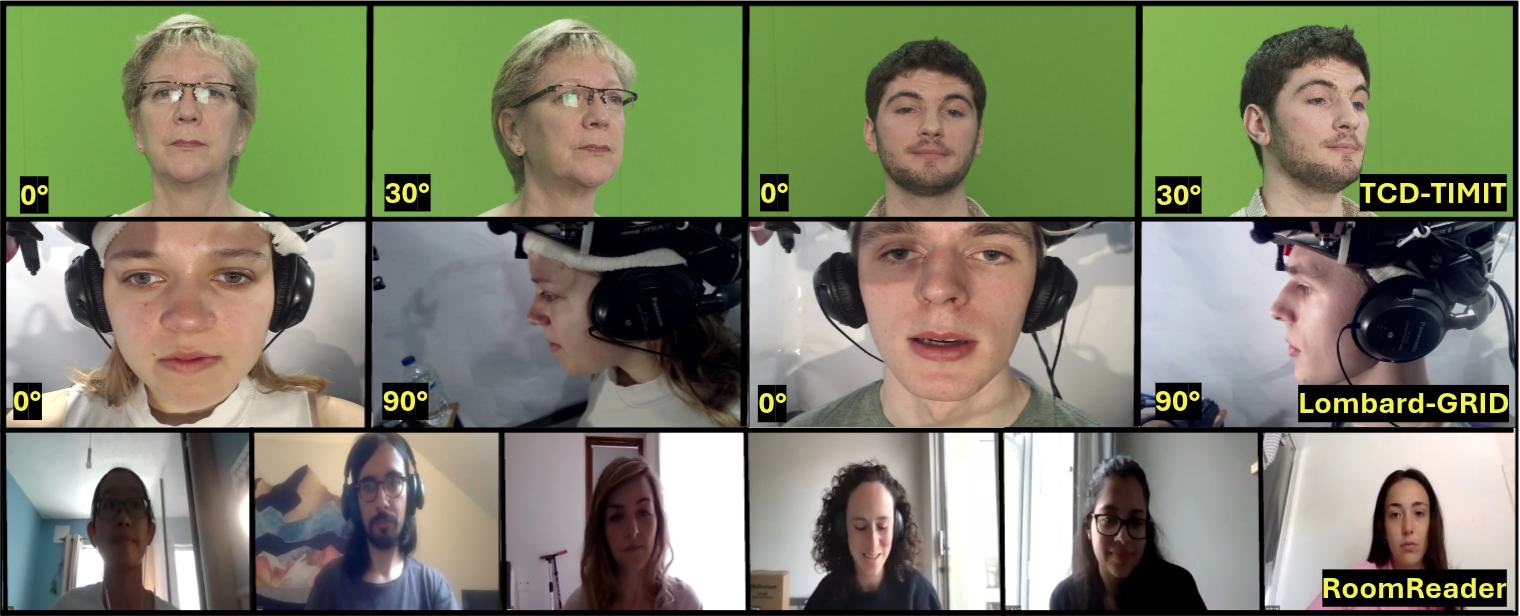}
\caption{Examples from the datasets used in this study. First row: professional lipspeaker and volunteer from TCD-TIMIT at 0$^\circ$ and 30$^\circ$ angles. Second row: LombardGrid speakers at 0$^\circ$ and 90$^\circ$ views. Third row: RoomReader Zoom conversations. GRID, LRS2, and LRS3 omitted due to space constraints.}
\label{fig:dataset_compare}
\vspace{-1.5em}
\end{figure}

Figure \ref{fig:dataset_compare} presents examples from the TCD-TIMIT, LombardGrid, and RoomReader datasets. For datasets other than LRS2 and LRS3, which use predefined test sets, we use the entire corpus for inference.

\subsection{Preprocessing Pipeline}
A practical barrier to multi-condition AVSR evaluation is not the absence of suitable datasets, but format incompatibility. Adapting new and differently formatted datasets to widely used frameworks like AV-HuBERT and Auto-AVSR (which serve as the foundation for most modern AVSR systems) requires substantial, undocumented engineering effort. As a result, researchers often avoid this bottleneck by evaluating only on standard benchmarks like LRS2 and LRS3, whose test sets are less than 1 hour. To address this, we release a preprocessing pipeline that converts GRID, LombardGrid, TCD-TIMIT, and RoomReader into formats directly compatible with both frameworks. This pipeline requires no modification to either codebase, allowing users to seamlessly integrate these datasets into existing research workflows.

The pipeline applies a consistent set of preprocessing steps across all datasets. First, faces are detected and mouth regions of interest (ROIs) are extracted using RetinaFace \cite{RF} with 68-point facial landmarks, producing 96$\times$96 pixel crops. Audio is standardized by downsampling to 16 kHz mono, while transcripts are normalized by removing punctuation and converting text to lowercase. For GRID dataset \cite{Grid}, which lacks explicit transcripts, we generate them by mapping them to their fixed six-word vocabulary pattern. For RoomReader \cite{reverdy2022roomreader}, although disfluency markers (\$ and \#) are removed, we explicitly retain backchannels and filled pauses (e.g., "yeah," "hmm") to preserve the natural flow of spontaneous conversation. Retaining these elements directly reflects real-world use cases and significantly increases the challenge of the benchmark. Finally, dataset manifests are generated in formats compatible with AV-HuBERT \cite{avhubert} and Auto-AVSR \cite{autoavsr}. The newly revamped RoomReader dataset produced by this pipeline is referred to as RoomReader-AV in this work. The complete pipeline, along with comprehensive documentation detailing the data cleaning steps required for each dataset, is publicly available on our GitHub\footnote{https://github.com/rishabhjain16/lipreading-data-guide}.

\section{AVSR Models}
We evaluate three state-of-the-art AVSR models using their official pretrained checkpoints. All decoding settings follow each model's official implementation and were held constant across datasets and modalities. Performance is measured in WER across VO, AO, and AV settings.

\textbf{Auto-AVSR} \cite{autoavsr} is trained on 3,448 hours of AV data (LRW \cite{lrw}, LRS2, LRS3, VoxCeleb2 \cite{vc2}, AVSpeech \cite{AVSpeech}) using a ResNet-18 visual frontend, Conformer encoder \cite{conformer}, and hybrid CTC/attention decoder \cite{watanabe2017hybrid}. It represents the large-scale \textbf{supervised} approach. For the scale analysis in Section \ref{sec:scale}, we also evaluate a checkpoint trained on 1,759 hours (LRS3 + VoxCeleb2), which matches AV-HuBERT's and Llama-AVSR's pretraining data volume and allows for a fairer comparison.

\textbf{AV-HuBERT }Large \cite{avhubert} uses masked multimodal prediction for \textbf{self-supervised} pretraining \cite{hsu2021hubert} using VoxCeleb2 and LRS3 datasets, followed by fine-tuning on LRS3. It processes multi-stream AV input with a shared Transformer. The noise-augmented checkpoint is used throughout this work.

\textbf{Llama-AVSR } \cite{llama_avsr} combines pretrained AV-HuBERT visual features with Whisper \cite{whisper} acoustic features and uses a Llama3.1-8B-based \cite{llama3} \textbf{LLM decoder}. It is finetuned with a LoRA \cite{hu2022lora} adapter on the LRS3 and VoxCeleb2 datasets.
\section{Experiment and Results}

\subsection{Large-Scale Lip-Reading Corpora: LRS2 and LRS3}
\begin{table}[t]
\centering
\caption{WER(\%) comparison on the LRS2 and LRS3 datasets across different modalities}
\label{tab:lrs_results}

\renewcommand{\arraystretch}{1.15}

\begin{tabular}{|l|c|c|c|c|c|c|}
\hline

\rowcolor{orange!20}
 & \multicolumn{3}{c|}{\textbf{LRS2}} & \multicolumn{3}{c|}{\textbf{LRS3}} \\
\cline{2-7}

\rowcolor{orange!20}
\textbf{Model} & \textbf{VO $\downarrow$} 
& \textbf{AO $\downarrow$} 
& \textbf{AV $\downarrow$} 
& \textbf{VO $\downarrow$} 
& \textbf{AO $\downarrow$} 
& \textbf{AV $\downarrow$} \\
\hline

Auto-AVSR  & 14.65 & 1.46 & 1.76 & 19.10 & 1.00 & 0.90 \\
AV-HuBERT  & 38.00 & 8.26 & 7.25 & 28.69 & 1.95 & 1.47 \\
Llama-AVSR & 41.57 & 5.07 & 4.58 & 26.20 & 0.74 & 0.79 \\
\hline

\end{tabular}

\vspace{1.5mm}

{\footnotesize
Results are reported using the best checkpoint released by the authors of each model. $\downarrow$: Lower is better.
}
\vspace{-1em}
\end{table}

Table \ref{tab:lrs_results} shows the AVSR results on LRS2 and LRS3. LRS3 is the shared training dataset across all three models. LRS2 is in-domain for Auto-AVSR but unseen for AV-HuBERT and Llama-AVSR. AV performance is at, or near, AO for all models on LRS3, confirming that visual fusion works when test conditions match training. Llama-AVSR achieves the best LRS3 AV (0.79\%) and AO (0.74\%), with Auto-AVSR close behind (0.90\% AV). On LRS2, Auto-AVSR achieves the strongest VO due to direct LRS2 training; however, for the models where LRS2 is unseen, AV-HuBERT outperforms Llama-AVSR in the VO setting (38.00\% vs. 41.57\%), showing stronger visual generalization. Even so, VO remains dramatically weaker than AO and AV for all models. 

\subsection{Structured Read Speech: GRID}

The GRID corpus consists of fixed six-word command grammar (e.g., \textit{place green at B 4 now}), which is entirely absent from all three models' training data, creating a domain mismatch in both vocabulary and sentence structure. As seen in Table \ref{tab:grid}, VO performance collapses for all models, reaching 66.53\% for Auto-AVSR, 83.80\% for AV-HuBERT, and exceeding 116\% for Llama-AVSR. More critically, AV performance is worse than AO for Auto-AVSR (-8.90\%) and substantially worse for Llama-AVSR (-41.88\%), while AV-HuBERT shows no change. Overall, the visual stream contributes little benefit under this domain shift and in most cases actively degrades recognition performance. Auto-AVSR achieves the best generalization across all modalities, reflecting its more diverse training data.

\begin{table}[t]
\centering
\caption{Benchmarking WER (\%) Results on GRID Dataset.}
\label{tab:grid}
\renewcommand{\arraystretch}{1.15}
\begin{tabular}{|l|c|c|c|c|}
\hline
\rowcolor{purple!20}
\textbf{Model} & \textbf{VO $\downarrow$} & \textbf{AO $\downarrow$} & \textbf{AV $\downarrow$} & \textbf{$\Delta$ (AO--AV)} \\
\hline
Auto-AVSR  & 66.53  & 12.57 & 21.47 & -8.90 \\
AV-HuBERT  & 83.80  & 42.42 & 42.98 & -0.56 \\
Llama-AVSR & 116.39 & 42.26 & 84.14 & -41.88 \\
\hline
\end{tabular}

\vspace{1mm}

{\footnotesize
$\Delta$ (AO--AV) denotes the difference between AO and AV WER.  Negative values indicate degradation from AV fusion. $\downarrow$: Lower is better. 
}
\vspace{-0.8em}

\end{table}
\begin{table}[t]
\centering
\caption{Error examples for AV-HuBERT on GRID dataset.}
\label{tab:ao_vo_errors}
\renewcommand{\arraystretch}{1.2}
\setlength{\tabcolsep}{8pt}
\small

\resizebox{\linewidth}{!}{%
\begin{tabular}{|l|l|l|}
\hline
\rowcolor{purple!20}
\textbf{Reference} & \textbf{AO} & \textbf{VO} \\
\hline
place blue in b 7 please  & place blew and be 7 please & but it's blue and pizzaval bees \\
lay blue with r 2 please  & lay blue with are 2 please  & light blewer beer 3 song \\
place white in i 6 now    & please wide in icex now     & yeah i'm here's why i sit down \\
lay white in c 1 again    & lay white in sea 1 again    & please wine i'll see what \\
\hline
\end{tabular}%
}

\vspace{1mm}
{\scriptsize
*AV-HuBERT AV results are omitted as they mirror the AO results.
}
\vspace{-1.5em}
\end{table}
The differences between the AO and VO errors in Table \ref{tab:ao_vo_errors} illustrate the contrasting failure modes of acoustic versus visual recognition on this structural grammar. The AO model relies entirely on sound, so its errors are primarily homophones. It captures the correct phonemic structure but outputs alternative spellings, such as mistaking the letters and numbers \textit{b 7} for \textit{be 7}, \textit{r 2} for \textit{are 2}, or \textit{c 1} for \textit{sea 1}. Conversely, the VO model relies entirely on visual cues, so it struggles with words that look identical on the lips (homophenes). To compensate for visual ambiguity, the model hallucinates phrases to satisfy its language decoder. This causes it to alter the syntax (turning \textit{place blue} into \textit{but it's blue}), drop articulatory cues (\textit{white} becoming \textit{why}), or introduce completely unrelated vocabulary (\textit{c 1 again} becoming \textit{wine i'll see what}).

\subsection{Lombard Speech Under Noise: LombardGrid}
\label{sec:lombardgrid}

\begin{table}[t]
\centering
\caption{Benchmarking WER (\%) performance on LombardGrid}
\label{tab:lombardgrid}
\renewcommand{\arraystretch}{1.15}
\begin{tabular}{|l|c|c|c|c|}
\hline
\rowcolor{red!25}
\textbf{Model} & \textbf{VO $\downarrow$} & \textbf{AO $\downarrow$} & \textbf{AV $\downarrow$} & \textbf{$\Delta$ (AO--AV)} \\
\hline
Auto-AVSR  & 78.69  & 14.02 & 11.73 & +2.29 \\
AV-HuBERT  & 94.80  & 34.04 & 33.64 & +0.40 \\
Llama-AVSR & 145.11 & 36.41 & 39.20 & -2.79 \\
\hline
\end{tabular}

\vspace{1mm}

{\footnotesize
$\downarrow$: Lower is better. $\Delta$ (AO--AV) denotes the difference between AO and AV WER. Positive values indicate improvement from audiovisual fusion, while negative values indicate degradation. 
}
\vspace{-1.5em}
\end{table}

LombardGrid represents the only out-of-domain scenario where AV consistently improves over AO, as shown in Table~\ref{tab:lombardgrid}. Results include the entire dataset, combining both the frontal and profile viewing conditions. While the recorded audio remains clean, speakers heard 80 dB noise through headphones to induce the Lombard effect \cite{4749459, lindblom1990explaining}. This causes speakers to naturally hyper-articulate and exaggerate their facial movements, making the dataset well-suited for demonstrating the benefits of visual fusion. In this scenario, Auto-AVSR benefits clearly from visual information, with AV WER (11.73\%) improving over AO (14.02\%). AV-HuBERT shows a small but consistent gain as well. In contrast, Llama-AVSR does not follow this trend, with AV performing slightly worse than AO. Furthermore, Llama-AVSR occasionally produces nonsensical outputs, such as severe repetitions and incoherent transcriptions, which is a characteristic failure mode of LLM-based decoders prone to hallucination when facing out-of-domain distribution shifts. Notably, Auto-AVSR's +2.29\% AV improvement directly reverses its -8.90\% degradation on the identical grammar of the standard GRID corpus (Table \ref{tab:grid}), demonstrating that hyper-articulated visual cues can compensate for severe syntactic mismatch.

\begin{table}[t]
\centering
\caption{WER(\%) performance on LombardGrid under frontal (0$^\circ$) and side-profile (90$^\circ$) views.}
\label{tab:lombardgrid_view}

\renewcommand{\arraystretch}{1.15}
\setlength{\tabcolsep}{6pt}

\begin{tabular}{|l|c|c|c|c|c|c|}
\hline

\rowcolor{red!25}
& \multicolumn{2}{c|}{\textbf{VO $\downarrow$}} 
& \multicolumn{2}{c|}{\textbf{AO $\downarrow$}} 
& \multicolumn{2}{c|}{\textbf{AV $\downarrow$}} \\
\cline{2-7}
\rowcolor{red!25}
\textbf{Model} & \textbf{0$^\circ$} & \textbf{90$^\circ$}
& \textbf{0$^\circ$} & \textbf{90$^\circ$}
& \textbf{0$^\circ$} & \textbf{90$^\circ$} \\
\hline

Auto-AVSR  & 64.88  & 92.50  & 14.03 & 14.03 & 11.59 & 11.87 \\
AV-HuBERT  & 87.06  & 102.53 & 34.04 & 34.04 & 33.11 & 34.17 \\
Llama-AVSR & 111.01 & 180.62 & 36.41 & 36.41 & 38.72 & 39.68 \\
\hline

\end{tabular}

\vspace{1mm}

{\footnotesize
$\downarrow$: Lower is better. $0^\circ$ denotes frontal (head-on) view, while $90^\circ$ denotes profile (side-view) condition. 
}
\vspace{-2em}
\end{table}

VO performance degrades significantly at the 90$^\circ$ (profile) view for all models (see Table~\ref{tab:lombardgrid_view}). Auto-AVSR WER increases from 64.88\% to 92.50\%, AV-HuBERT from 87.06\% to 102.53\%, and Llama-AVSR from 111.01\% to 180.62\%, indicating a strong sensitivity of visual recognition to head pose. In contrast, AO performance remains unchanged across views, as expected, since it does not depend on visual input. AV performance shows only minor changes, with Auto-AVSR moving from 11.59\% to 11.87\% and AV-HuBERT from 33.11\% to 34.17\%. This suggests that AV robustness under profile views is not due to reliable visual recognition at extreme poses, but rather due to the dominance of the audio stream in fusion when visual information becomes unreliable, with the exception of Auto-AVSR (90° AV 11.87\% vs. AO 14.03\%).

\subsection{Camera Angle and Speaker Articulation: TCD-TIMIT}
TCD-TIMIT enables us to independently examine the effects of \textbf{camera angle}, using its 0° and 30° recordings, and \textbf{articulation quality}, by comparing professional lipspeakers with everyday volunteers (non-professional speakers).

\begin{table}[t]
\centering
\caption{WER (\%) on TCD-TIMIT across viewing angles (0$^\circ$ and 30$^\circ$).}
\label{tab:timit_angle}
\renewcommand{\arraystretch}{1.15}
\setlength{\tabcolsep}{3.5pt}

\begin{tabular}{|l|c|c|c|c|c|c|c|}
\hline
\rowcolor{green!30!blue!20}
 & & \multicolumn{3}{c|}{\textbf{Lipspeakers}} & \multicolumn{3}{c|}{\textbf{Volunteers}} \\
\cline{3-8}
\rowcolor{green!30!blue!20}
\textbf{Model} & \textbf{Mod.} & \textbf{0$^\circ \downarrow$} & \textbf{30$^\circ \downarrow$} & \textbf{$\Delta$} & \textbf{0$^\circ \downarrow$} & \textbf{30$^\circ \downarrow$} & \textbf{$\Delta$} \\
\hline
Auto-AVSR   & \multirow{3}{*}{VO} & 25.91 & 26.54 & 0.63 & 45.57 & 47.68 & 2.11 \\
AV-HuBERT   &                     & 38.73 & 38.76 & 0.03 & 63.73 & 65.15 & 1.42 \\
Llama-AVSR  &                     & 43.90 & 45.51 & 1.61 & 87.95 & 90.78 & 2.83 \\
\hline
Auto-AVSR   & \multirow{3}{*}{AO} & 4.14  & 4.16  & 0.02 & 6.59  & 6.45  & 0.14 \\
AV-HuBERT   &                     & 11.08 & 11.13 & 0.05 & 16.83 & 16.37 & 0.46 \\
Llama-AVSR  &                     & 5.54  & 5.60  & 0.06 & 8.47  & 7.88  & 0.59 \\
\hline
Auto-AVSR   & \multirow{3}{*}{AV} & 7.06  & 7.20  & 0.14 & 10.88 & 10.24 & 0.64 \\
AV-HuBERT   &                     & 11.89 & 11.82 & 0.07 & 18.24 & 17.85 & 0.39 \\
Llama-AVSR  &                     & 10.98 & 10.99 & 0.01 & 15.25 & 14.64 & 0.61 \\
\hline
\end{tabular}

\vspace{1mm}
{\footnotesize
$\downarrow$: lower is better. Mod. = Modalities. 0$^\circ$ = frontal view; 30$^\circ$ = semi-profile view. 
$\Delta = |0^\circ - 30^\circ|$ (absolute performance gap between angles). 
}
\vspace{-1em}
\end{table}

\begin{table}[t]
\centering
\caption{WER (\%) in TCD-TIMIT across varying modality for Lipspeakers, Volunteers, and Matched Volunteers.}
\label{tab:matched_volunteers}
\renewcommand{\arraystretch}{1.15}
\setlength{\tabcolsep}{5pt}
\begin{tabular}{|l|c|c|c|c|c|}
\hline
\rowcolor{green!30!blue!20}
\textbf{Model} & \textbf{Mod.} & \textbf{LS $\downarrow$} & \textbf{V $\downarrow$} & \textbf{MV $\downarrow$} & \textbf{$\Delta$ (LS$-$MV)} \\
\hline
Auto-AVSR   & \multirow{3}{*}{VO} & 26.22 & 46.63 & 43.15 & -16.93 \\
AV-HuBERT   &                     & 38.75 & 64.44 & 60.39 & -21.64 \\
Llama-AVSR  &                     & 43.90 & 89.46 & 87.95 & -44.05 \\
\hline
Auto-AVSR   & \multirow{3}{*}{AO} & 4.15  & 6.72  & 6.08  & -1.93  \\
AV-HuBERT   &                     & 11.08 & 16.60 & 16.83 & -5.75  \\
Llama-AVSR  &                     & 5.54  & 8.16  & 8.47  & -2.93  \\
\hline
Auto-AVSR   & \multirow{3}{*}{AV} & 7.13  & 10.56 & 9.49  & -2.36  \\
AV-HuBERT   &                     & 11.89 & 18.04 & 18.24 & -6.35  \\
Llama-AVSR  &                     & 10.98 & 14.98 & 15.25 & -4.27  \\
\hline
\end{tabular}

\vspace{1mm}
{\footnotesize
$\downarrow$: lower is better. Mod. = Modalities. LS = Lipspeakers, V = Volunteers, MV = Matched Volunteers. $\Delta$ (LS$-$MV) represents the performance gap between LS and MV. 
}
\vspace{-1.5em}
\end{table}

\subsubsection{Effect of Camera Angle in TCD-TIMIT}

We evaluate all models under two viewing conditions, frontal (0$^\circ$) and semi-side profile (30$^\circ$), across Lipspeakers and Volunteers as presented in Table~\ref{tab:timit_angle}. Across all configurations, the change in WER between 0$^\circ$ and 30$^\circ$ remains small, consistently below 3\% absolute difference. AO performance is unchanged across viewing angles, as expected since it does not depend on visual input. Similarly, AV performance shows no meaningful variation with pose change in this moderate range. This suggests that small frontal offsets are within the robustness range of current visual encoders. When combined with the LombardGrid 90$^\circ$ results in Section \ref{sec:lombardgrid}, a consistent pattern emerges. While moderate pose changes have negligible impact, extreme profile views lead to substantial degradation in visual recognition. In those cases, AV performance is largely maintained through reliance on the audio stream rather than true visual pose invariance. This aligns with findings by Lan et al. \cite{view}, who showed that for computer lip-reading systems, moderate angles preserve the visibility of important articulatory gestures.

\subsubsection{Speaker Articulation in TCD-TIMIT}
All three professional lipspeakers in TCD-TIMIT are female. To enable a fair comparison with the volunteer group, we define a matched volunteers (MV) subset of female volunteers with identical sentences, durations, and utterance counts to the lipspeakers. This subset helps us isolate the effect of articulation, as professional lipspeakers are trained to produce clear visible lip movements while volunteers are everyday speakers with no such training.

The articulation gap between Lipspeakers and Matched Volunteers is large and highly consistent across all settings (Table~\ref{tab:matched_volunteers}). The precise articulation of the professional lipspeakers is directly reflected in the VO performance. For example, at 0$^\circ$, Auto-AVSR achieves a 25.91\% WER on lipspeakers compared to 45.57\% on the matched volunteers. AV-HuBERT and Llama-AVSR exhibit similar, substantial performance gaps between the two groups, confirming that professional articulation significantly aids visual recognition. Volunteers remain significantly harder to recognize in the VO setting, with performance drops ranging from approximately 17\% to 44\% depending on the model. In contrast, AO performance shows only minor degradation, typically between 1\% and 6\%, indicating that the acoustic signal is similarly clean across both speaker groups. However, this visual advantage does not translate to improved AV performance. Across all models, AV performance is consistently worse than AO performance. The TCD-TIMIT audio recordings are exceptionally clean and AO performance is already highly accurate. In this scenario, fusing the less reliable visual modality actively degrades the near-perfect audio stream, resulting in a negative AV fusion effect regardless of the speaker's articulation quality.

\subsection{Spontaneous Conversational Speech: RoomReader-AV}
\label{sec:rr_av}

\begin{table}[t]
\centering
\caption{WER (\%) on RoomReader-AV}
\label{tab:roomreader}
\renewcommand{\arraystretch}{1.15}
\setlength{\tabcolsep}{6pt}

\begin{tabular}{|l|c|c|c|c|}
\hline
\rowcolor{blue!10!green!15}
\textbf{Model} & \textbf{VO $\downarrow$} & \textbf{AO $\downarrow$} & \textbf{AV $\downarrow$} & \textbf{$\Delta$ (AO$-$AV)} \\
\hline
Auto-AVSR   & 110.07 & 25.40 & 26.39  & -0.99  \\
\hline
AV-HuBERT   & 92.54  & 36.18 & 34.44  & +1.74   \\
\hline
Llama-AVSR  & 311.48 & 88.01 & 128.54 & -40.53 \\
\hline
\end{tabular}

\vspace{1mm}
{\footnotesize
$\downarrow$: lower is better. $\Delta$ (AO--AV) = AO--AV WER difference; positive = improvement from AV fusion, negative = degradation from AV fusion.
}
\end{table}

\begin{figure}[t]
\centering
\includegraphics[width=1\linewidth]{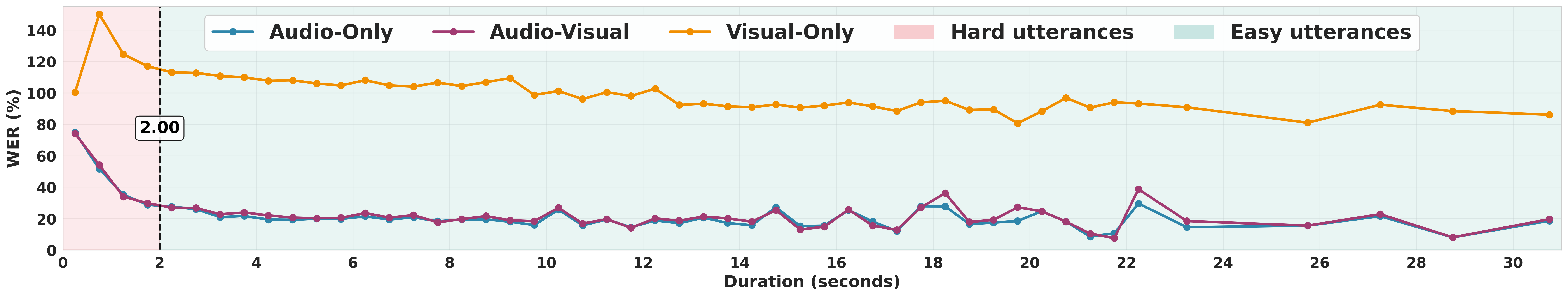}
\caption{WER (\%) of the Auto-AVSR model across utterance durations (in seconds) for the VO, AO, and AV modalities on RoomReader-AV.}
\label{fig:wer_roomreader}
\vspace{-1em}
\end{figure}

\begin{table}[t]
\centering
\caption{WER (\%) on RoomReader-AV Subsets: Easy and Hard}
\label{tab:easy_hard}
\renewcommand{\arraystretch}{1.15}
\setlength{\tabcolsep}{5pt}
\begin{tabular}{|l|c|c|c|c|c|c|}
\hline
\rowcolor{blue!10!green!15}
 & \multicolumn{3}{c|}{\textbf{RR-AV\_Easy  ($\geq$ 2s)}} & \multicolumn{3}{c|}{\textbf{RR-AV\_Hard  ($<$ 2s)}} \\
\cline{2-7}
\rowcolor{blue!10!green!15}
\textbf{Model} & \textbf{VO $\downarrow$} & \textbf{AO $\downarrow$} & \textbf{AV $\downarrow$} & \textbf{VO $\downarrow$} & \textbf{AO $\downarrow$} & \textbf{AV $\downarrow$} \\
\hline
Auto-AVSR & 105.35 & \textbf{20.60} & \textbf{21.72} & 126.95 & \textbf{42.58} & \textbf{43.14} \\
AV-HuBERT & \textbf{89.72}  & 28.61 & 26.90 & \textbf{96.65}  & 53.75 & 52.24 \\
\hline
\end{tabular}

\vspace{1mm}
{\footnotesize $\downarrow$: lower is better. Llama-AVSR is excluded due to high error rates.}
\vspace{-2em}
\end{table}

RoomReader-AV is the most challenging subset in this evaluation, with no in-domain exposure for any model. For our evaluation, we use the individual audio stream, featuring each speaker's isolated audio. Several patterns emerge from the results in Table \ref{tab:roomreader}. VO collapses for all models. AO performance is also challenged but remains viable for Auto-AVSR (25.40\%) and AV-HuBERT (36.18\%), whereas Llama-AVSR struggles significantly with the AO condition (88.01\%). Auto-AVSR AV closely follows its AO performance, indicating it degrades without overly relying on the visual stream. AV-HuBERT is the only model to show an AV improvement, suggesting it can extract some complementary visual cues even in adverse conditions. Llama-AVSR suffers a severe AV penalty, with its WER increasing by over 40 compared to its AO result. Llama-AVSR’s extreme VO error rate of 311.48\% reflects severe hallucinations. For short backchannel inputs like "yeah" or "hmm," the model generates long, unrelated sequences, as pretrained language model knowledge overrides both visual and acoustic evidence. This represents a fundamentally different failure mode from other architectures.

\textbf{Why is VO much worse than AO on RoomReader-AV?} 
Both modalities struggle, but audio degrades less significantly. Several factors explain the visual collapse. First, spontaneous meeting speech is informal, with casual vocabulary, fillers, incomplete sentences, and frequent topic shifts, which differs significantly from the scripted or semi-scripted speech the models were trained on. Second, Zoom participants often look away from the camera, either to their screens or while gesturing, leading to suboptimal mouth ROI quality. Third, many RoomReader-AV turns are short ("yeah," "okay," "right"), offering minimal visual context, even for human lipreaders. Finally, the AVSR models were trained primarily on broadcast-style speech, which differs significantly from the more dynamic facial movements seen in meeting scenarios.

To further investigate the impact of these short turns, in Figure~\ref{fig:wer_roomreader}, we plot the Auto-AVSR model WER for VO, AO, and AV modes against utterance duration. The plot clearly shows that WERs are much higher for short clips and drop rapidly as clips get longer. This observation is strongly supported by recent research from Djilali et al. \cite{djilali2024vsr}, which highlights that utterances shorter than 2 seconds present an inherent recognition challenge due to limited temporal context. Driven by this observation, we partition the RoomReader-AV into two difficulty subsets using this established 2-second threshold. Clips shorter than 2 seconds form the hard subset: \textbf{RR-AV\_Hard (6,752 utterances, 1.6 hours)}, while clips of 2 seconds or longer form the easy subset: \textbf{RR-AV\_Easy (3,572 utterances, 4.9 hours)}. Clips in the RR-AV\_Hard average just 2.3 words, compared to 15.9 words in the RR-AV\_Easy.

Table~\ref{tab:easy_hard} benchmarks the models across these subsets. The Hard subset presents a severe recognition challenge across all configurations. Most notably, AO WER effectively doubles for both models when evaluated on the Hard subset. While the visual modality remains extremely challenging across both subsets, it exhibits a similarly pronounced degradation on the Hard subset. These subsets are intended to provide a more challenging evaluation benchmark for AVSR research, reducing over-reliance on controlled datasets such as LRS2 and LRS3 where models have largely saturated performance.

\subsection{Scaling and Generalization in Visual Speech Recognition}
\label{sec:scale}
We compare models trained on varying data scales to assess the impact of dataset size and architecture on VO WER. To ensure a fair baseline, we include Auto-AVSR trained on 1,759h, which uses the exact same training data as AV-HuBERT and Llama-AVSR. This setup isolates how well visual features generalize across different datasets and training conditions. As shown in Table \ref{tab:scale}, data scale heavily impacts VSR performance. Increasing Auto-AVSR's training data from 1.7k to 3.4k hours leads to consistent, significant reductions in VO WER across most datasets. This improvement plateaus for RoomReader-AV, where performance remains almost unchanged despite doubling the training data. This indicates that while scaling data helps in structured, in-domain scenarios, generalizing to diverse, real-world environments is a fundamental problem, not merely a matter of scale. As doubling broadcast data fails to resolve this out-of-domain visual bottleneck, achieving true generalization will likely require training on conversational "in-the-wild" data or explicitly decoupling visual representation learning from AV fusion. At the 1.7k-hour scale, the supervised Auto-AVSR demonstrates stronger out-of-domain robustness compared to AV-HuBERT, with RoomReader being the only exception, whereas Llama-AVSR only performed well with seen LRS3.

\begin{table}[t]
\centering
\caption{Cross-dataset performance WER (\%) across models and training scales for Video-Only Mode.}
\label{tab:scale}
\renewcommand{\arraystretch}{1.15}
\setlength{\tabcolsep}{4.5pt}
\begin{tabular}{|l|c|c|c|c|}
\hline
\rowcolor{yellow!20}
\textbf{Dataset} & \textbf{A-3.4k $\downarrow$} & \textbf{A-1.7k $\downarrow$} & \textbf{AVH $\downarrow$} & \textbf{Llama-AVSR $\downarrow$} \\
\hline
LRS2             & \textbf{14.65}  & 35.05   & 38.00 & 41.57  \\
LRS3             & \textbf{19.10}  & 31.14   & 34.04 & 26.20  \\
GRID             & \textbf{66.53}  & 83.92   & 88.53 & 116.39 \\
LombardGrid      & \textbf{78.69}  & 94.69   & 94.80 & 145.11 \\
TCD-TIMIT      & \textbf{43.18}  & 59.09   & 60.11 & 81.91  \\
RoomReader-AV     & 110.07 & 113.20  & \textbf{92.54} & 311.48 \\
\hline
\end{tabular}

\vspace{1mm}
{\footnotesize
A-3.4k = Auto-AVSR (3,448h); A-1.7k = Auto-AVSR (1,759h); AVH = AV-HuBERT. The TCD-TIMIT value is the averaged score across all Lipspeakers and Volunteers at both $0^\circ$ and $30^\circ$ angles. $\downarrow$: lower is better.
}
\vspace{-1em}
\end{table}

\section{Discussion}

\textbf{VO does not generalize outside its training domain:} Across all tested conditions, VO yields low error rates for in-domain (19–44\% WER) and completely collapses on every out-of-domain dataset (66–313\% WER). Visual encoders heavily overfit to the specific articulation and vocabulary of professional TED-style broadcast speakers, failing to transfer to other datasets and conditions. A model can achieve a 19\% VO WER on LRS3, while failing completely on every other condition; therefore, this performance does not equate to generalized visual speech understanding.

\textbf{AV fusion is beneficial in two situations:} 
AV mainly improves upon AO in-domain (LRS2, LRS3) and when lip movements are hyper-articulated (LombardGrid). In all other conditions, AV offers little benefit and can significantly worsen AO performance. Current fusion mechanisms blindly trust visual features regardless of quality, injecting noise into the audio pathway rather than down-weighting or ignoring them. The severe RoomReader AV failure for Llama-AVSR (+40 WER) provides clear evidence of this, while Auto-AVSR's stability stems from its large-scale supervised training yielding a more robust visual encoder.

\textbf{Camera pose tolerance is audio dominance, not visual robustness:} The LombardGrid angle comparison shows that AV tolerance at a 90$^\circ$ profile view comes mainly from audio fallback. When VO performance collapses at this extreme angle, the AV system relies on audio, maintaining near-identical performance across both views. TCD-TIMIT confirms this: a 30$^\circ$ offset is tolerated because it remains within the visual encoder's frontal range, whereas a 90° angle renders visual features uninformative and the model falls back to audio.

\textbf{Speaker articulation matters much more than camera position:} The 17 to 44\% error gap between professional lipspeakers and everyday volunteers is much larger than the minimal 2\% change caused by a 30$^\circ$ camera offset. Since benchmark tests use trained professionals, they do not reflect how normal people talk. Therefore, any claims about real-world AVSR performance must account for this major difference in speaking style.

\textbf{LLM-based decoders: In-domain gains vs. deployment costs:} While Llama-AVSR achieves superior in-domain results, its LLM decoder introduces severe out-of-domain instability, often producing hallucinated, repetitive outputs on short spontaneous inputs. Beyond these accuracy drops, the architecture imposes a massive computational penalty. Since Llama-AVSR shares Auto-AVSR's codebase, this hardware comparison is direct. Llama-AVSR requires 17.91 GB of GPU memory for AV inference compared to the 2.29 GB needed by Auto-AVSR. Furthermore, while model training costs are outside our current scope, the resource disparity in that phase would be significantly greater. This extreme overhead, coupled with slow autoregressive generation, renders the model largely unsuitable for low-latency, real-time applications.

\section{Conclusion and Future Work}
This paper presents a unified multi-condition evaluation of AVSR across six diverse datasets, assessing supervised, self-supervised, and LLM-based architectures. Our work demonstrates that current AVSR progress is narrow in scope. Regardless of architecture or training scale, visual-only models fail to generalize outside of broadcast speech, and apparent visual robustness to extreme camera angles is largely driven by acoustic fallback. Additionally, the use of LLMs introduces new challenges to domain shifts. Evaluation on standard datasets like LRS2 and LRS3 is insufficient to characterize real-world system generalization. Thus, to address this gap and drive the field toward genuine robustness, we introduce RoomReader-AV as a new benchmark for spontaneous multi-party conversational AVSR. Furthermore, we release a preprocessing pipeline that converts this novel dataset, alongside GRID, LombardGrid, and TCD-TIMIT, into formats directly compatible with existing AVSR frameworks, making comprehensive multi-condition evaluation highly accessible and extension to newer ASR/VSR/AVSR systems straightforward.

\section*{AI-Generated Content Disclosure}
During the preparation of this work, Perplexity (Claude Sonnet 4.6) was used exclusively for minor English grammar corrections and improving the clarity of the written text.

\bibliographystyle{IEEEtran}
\bibliography{my_bib}

\end{document}